# Strong $Eu^{3+}$ luminescence in $La_{1-x-y}Er_{x/2}Eu_{x/2}Ca_yVO_4$ nanocrystals: the result of co-doping optimization

*O. Chukova[1]*, S.A. Nedilko[1], S.G. Nedilko[1], T. Voitenko[1], A. Slepets[2], M. Androulidaki[3], A. Papadopoulos[3], E. Stratakis[3], W. Paszkowicz[4]*

1 – Taras Shevchenko National University of Kyiv, Volodymyrska Str., 64/13, Kyiv 01601, Ukraine

2 – Bogomolets National Medical University, T. Shevchenko blvd., 13, Kyiv 01601, Ukraine

3 – Institute of Electronic Structure & Laser (IESL) of Foundation for Research & Technology Hellas (FORTH), Heraklion 711 10, Crete, Greece

4 – Institute of Physics, Polish Academy of Sciences, Warsaw, Poland

**Abstract**

Co-doped with $Ca^{2+}$ and $Er^{3+}$ ions $LaVO_4$:Eu crystalline nanoparticles are synthesized and investigated with a goal to clarify the mechanisms of $Ca^{2+}$ and $Er^{3+}$ impurities effects on $Eu^{3+}$ ions luminescence and to find compositions with enhanced luminescence intensity. The XRD analysis reveals dependence of crystal structure on dopants concentration: monoclinic crystal phase is observed for low dopant concentrations and content of tetragonal crystal phase increases with dopant concentrations increase. The SEM investigation reveals formation of nanoparticles with two types of shapes. Photoluminescence spectra consist of lines caused by *f-f* transitions in the $Er^{3+}$ and $Eu^{3+}$ ions. It has been shown that dependence of the $Eu^{3+}$ ions luminescence intensity on the $Er^{3+}$ and $Ca^{2+}$ concentrations is caused by cumulative effects of dopants on crystal lattice structure, on defects in the first coordination sphere of the $Eu^{3+}$ ions and on efficiency of excitation energy transfer. The maximal luminescence intensity is found for the $La_{0.8}Er_{0.05}Eu_{0.05}Ca_{0.1}VO_4$ composition. The corresponded intensity is as much as $19 \pm 2$ and $8 \pm 2$ times higher than luminescence intensity of the $La_{0.95}Eu_{0.05}VO_4$ and $La_{0.9}Eu_{0.05}Ca_{0.05}VO_4$ compounds, respectively. Conclusion is made that $Er^{3+}$ and $Ca^{2+}$ ions co-doping is a promising way to increase luminescence efficiency of the $Eu^{3+}$ ions in the $LaVO_4$ nanocrystals.

## 1. Introduction

The RE-containing orthovanadates of general formulae $LnVO_4$ (Ln = Y, Gd, …, Lu) are widely used for many decades and, at the same time, they are still under development as optical materials: phosphors, luminescent probes, markers, solid state and fiber laser hosts, amplifiers for fiber-optic communication, polarizers, etc. [1-8]. The modern solar energy revolution requires materials with improved characteristics to be used in solar concentrators, absorbing solar coating and incident light driven photocatalysis [9-12]. High thermal and radiation stabilities of oxides make them the main candidates for such applications [12-14]. And orthovanadates are considered and developed now for the noted applications [15-18]. In traditional applications of orthovanadates, head place is occupied by compounds based on yttrium orthovanadate ($YVO_4$), in particular, doped with europium [19, 20]. Recently, considerable attention has been paid also to the development of luminescent materials based on lanthanum orthovanadate, $LaVO_4$ [21 - 25]. This material is more widespread and its cost is much lower than $YVO_4$. However, it has a significant disadvantage: at the contrary to $YVO_4$, which crystallizes in a tetragonal crystal lattice of the zircon type, a stable modification of $LaVO_4$ is a monoclinic crystal lattice of the monazite type (m-$LaVO_4$). This feature leads to comparably low quantum yield and luminescence intensity of the impurity $Eu^{3+}$ ions in $LaVO_4$ matrix compared to the $Eu^{3+}$ in $YVO_4$ matrix [23, 26 - 28]. It has been observed that in orthovanadate crystal lattices, the excitation of the RE ions photoluminescence takes place mainly due to the excitation light absorption by the $VO_4^{3-}$ vanadate molecular groups [29 - 32]. After the absorption, the energy of the $VO_4^{3-}$ excited states non-radiatively migrates over the crystal lattice and then it is delivered to the RE ions. Since the distances between the $VO_4^{3-}$ groups in the m-$LaVO_4$ lattice (3.38 Å) is significantly larger than the corresponding distances in the $YVO_4$ lattice (3.26 Å) [24], the rate of energy transfer in the monoclinic $LaVO_4$ lattice is significantly lower, and therefore the excitation energy losses on the way to the RE luminescent centers are significantly higher. The situation can be improved by using a tetragonal modification of the lanthanum orthovanadate (t-$LaVO_4$), which is structurally similar to the $YVO_4$, but, unfortunately, it is metastable.

The tetragonal t-$LaVO_4$ phase can be obtained in the form of nanosized crystals, but it is still necessary to use suitable methods to stabilize this phase. To perform the directed synthesis of t-$LaVO_4$ phase a selection of various precursors, adding of oleic or ethylenediaminetetraacetic (EDTA) acids [23, 33] or other negatively charged ligands [26, 34], which coordinate the surface of the $LaVO_4$ nanoparticles, variation of pH or high pressure regimes, etc. are used [ 21, 28, 35].

The t-LaVO4 nanocrystals of synthesized by such methods have some disadvantages those are long-time storage instability and dependence of their structure and properties on temperature and pressure. This is caused because the stabilization of the t-phase is provided mainly by contribution of the surface, whereas bonds of the applied organic compounds with the surface are unstable [34, 36].

Dopant induced stabilization is another, but also widely used, way for directed stabilization of the crystal phase. This method is based on differences in the sizes of the lattice's cations and the stabilization impurity cations, which replace them. Namely, replacement of the $La^{3+}$ ions by smaller ions from a lanthanide (Ln) row causes a decrease in the number of oxygen ions from 9 to 8 in the $La^{3+}$ ion surrounding. This leads to the m-$LaVO_4$ → t-$LaVO_4$ transformation because the eight-coordinated $Ln^{3+}$ ion is a characteristic of the tetragonal $LnVO_4$ orthovanadate crystal lattice. Thus, incorporation of the luminescent $Ln^{3+}$ ions is a promising way for stabilization of the tetragonal phase. In this case, some impurity RE ions (Sm, Eu, Dy, etc.) not only stabilize the t-$LaVO_4$ phase, but also satisfy the expected luminescent properties [28, 37].

However, it is obvious that the situation when the same ions play a dual role limits the possibilities of the dopant induced stabilization approach.

Recently, in some published studies had been proposed to use multiple-doping to enhance the functionality of lanthanum orthovanadate doped with $Eu^{3+}$ ions. Thus, there are the papers [38, 39], where influence of ($Y^{3+}$, $Eu^{3+}$), ($Gd^{3+}$, $Eu^{3+}$), and ($Lu^{3+}$, $Eu^{3+}$) co-doping on structure and luminescence properties of $LaVO_4$ has been studied.

In this paper, we modify the approach considered in [38] and choose $Er^{3+}$ ions for the co-doping of the m-$LaVO_4$:Eu nanocrystals. The choice is done taking into account that $Er^{3+}$ ions are luminescent active in the visible range of light, and their luminescent characteristics in orthovanadate crystals are known [40, 41]. These reasons give possibility to use the $Er^{3+}$ dopants also as a luminescent probe to monitor the structural transformations of the m-$LaVO_4$:Eu nanocrystals. Moreover, in this work we introduce the $Ca^{2+}$ dopants in the m-$LaVO_4$:Eu composition. (Doping with alkali earth and some other two-charged cations (Ca, Sr, Ba, Zn) has been already used previously to enhance the intensity of the RE ions luminescence in the $LnVO_4$ nanocrystals and to achieve excitation effectiveness from near UV and violet spectral range [42–44].) Thus, we apply two the above-mentioned approaches in this paper. That is why, the purpose of the present work is to study a common effect of the $Er^{3+}$ and $Ca^{2+}$ ions on structural and optical properties of the $La_{1-x}Eu_xVO_4$ ($0 \leq x \leq 0.2$) crystalline nanoparticles in order to find compositions with enhanced intensity of the $Eu^{3+}$ ions luminescence and to clarify the mechanisms of concentration behavior of luminescence.

## 2. Experimental

The $La_{1-x-y}Er_{x/2}Eu_{x/2}Ca_yVO_4$ ($0 \leq x \leq 0.3$, $0 \leq y \leq 0.2$) powder nanoparticles were prepared by aqueous nitrate-citrate sol-gel synthesis. Detailed description of sol-gel synthesis procedure of the similar vanadates can be found in our previous papers [45, 46]. Solutions of lanthanum (III), europium (III), erbium (III) and calcium (II) nitrates, ammonium metavanadate, ammonium hydroxide, citric acid and nitric acid were used at synthesis process. The precipitate of reacted nitrate solutions was dissolved in a citric acid and concentrated by slow evaporation at 80-90 ºC before formation of a gel. Then a fine-grained powder was obtained from the gel using annealing at 680 ºC for 5 hours and homogenization in an agate mortar. The investigated in this paper concentration row of the samples is presented by the compositions $La_{0.9}Er_{0.025}Eu_{0.025}Ca_{0.05}VO_4$, $La_{0.8}Er_{0.05}Eu_{0.05}Ca_{0.1}VO_4$, $La_{0.65}Er_{0.1}Eu_{0.1}Ca_{0.15}VO_4$ and $La_{0.5}Er_{0.15}Eu_{0.15}Ca_{0.2}VO_4$ samples.

Phase compositions of the synthesized samples were determined using X-ray diffractometer Shimatzu 2000 ($Cu_{K\alpha}$-radiation with a Ni filter). The microstructure of the samples was studied with a scanning electron microscope (SEM) INCA X-max System from Oxford Instruments. Reflection spectroscopy of the samples was performed using Perkin Elmer Lambda 950 spectrometer. The powder samples were pressed in sample holder and then spectra of diffuse reflection were measured. In the used mode, the monochromator is placed before the samples and all light reflected from the powder samples are collected by photometric sphere. Luminescence spectra excited with 325 nm laser were registered using ACTON (500) monochromator with grating 150 grooves/mm (blaze @ 300 nm), slit on 50 micron and liquid $N_2$ - cooled CCD camera. Luminescence spectra excited with 405, 478 and 532 nm lasers or powerful Xenon lamp 325 nm were registered using DFS-12 monochromator with grating 600 grooves/mm, slit on 50 micron and FEU-79 photomultiplier [45].

## 3. Results

### 3.1. Crystal structure and phase compositions

X-Ray diffraction (XRD) patterns of the synthesized $La_{1-x-y}Er_{x/2}Eu_{x/2}Ca_yVO_4$ nanoparticles are shown in Fig. 1. For the $La_{0.9}Er_{0.025}Eu_{0.025}Ca_{0.05}VO_4$ sample only characteristic peaks of the monoclinic $LaVO_4$ structure were observed (according to the reference pattern, JCPDS PDF2 50-0367 $P_{21/n}$ space group). For the $La_{0.8}Er_{0.05}Eu_{0.05}Ca_{0.1}VO_4$ samples the traces of some characteristic peaks of the tetragonal $LaVO_4$ structure were also observed (JCPDS PDF2 32-0504 standard card, $I_{41/amd}$ space group). Content of the tetragonal $LaVO_4$ phase increases with increase of dopants concentrations. Thus, the $La_{0.65}Er_{0.1}Eu_{0.1}Ca_{0.15}VO_4$ sample is a two-phase mixture, and finally the highest doped $La_{0.5}Er_{0.15}Eu_{0.15}Ca_{0.2}VO_4$ sample is single phase that is tetragonal.

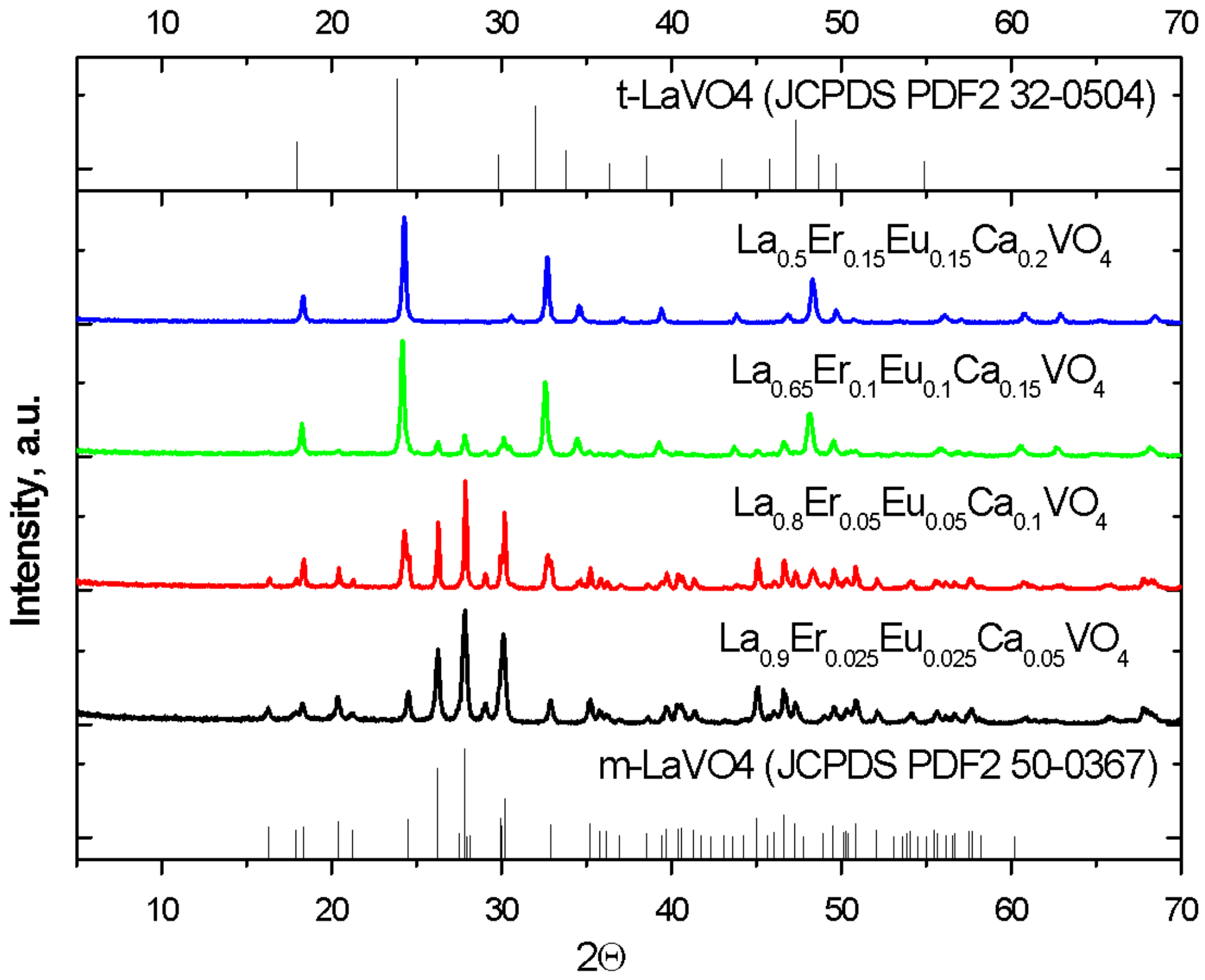

Fig. 1. The XRD patterns of the $La_{0.9}Er_{0.025}Eu_{0.025}Ca_{0.05}VO_4$ (1), $La_{0.8}Er_{0.05}Eu_{0.05}Ca_{0.1}VO_4$ (2), $La_{0.65}Er_{0.1}Eu_{0.1}Ca_{0.15}VO_4$ (3) and $La_{0.5}Er_{0.15}Eu_{0.15}Ca_{0.2}VO_4$ (4) nanoparticles

### 3.2. Morphology of the samples

SEM investigation of the synthesized samples revealed that synthesized samples consist of the nanoparticles with the sizes from 40 to 150 nm (Fig. 3). The $La_{0.9}Er_{0.025}Eu_{0.025}Ca_{0.05}VO_4$ and $La_{0.5}Er_{0.15}Eu_{0.15}Ca_{0.2}VO_4$ samples are characterized by good homogeneity of the nanoparticles sizes with 40 – 60 and 50 – 80 nm sizes, respectively (Fig. 3,a, d). The $La_{0.8}Er_{0.05}Eu_{0.05}Ca_{0.1}VO_4$ and $La_{0.65}Er_{0.1}Eu_{0.1}Ca_{0.15}VO_4$ samples contain nanoparticles of two types of sizes those are clearly seen in Fig. 3, b, c. There are smaller nanoparticles (40 – 60 nm) and bigger nanoparticles. The bigger-size nanoparticles have sizes up to 100 - 150 nm. For the $La_{0.8}Er_{0.05}Eu_{0.05}Ca_{0.1}VO_4$ sample, content of the smaller-size nanoparticles is essentially lower compared to the bigger-size ones (Fig. 3,b).

Comparing the results of SEM investigation with XRD analysis of crystal phase composition, we can suppose that formation of two types of nanoparticles is connected with presence of two crystal phases in the synthesized samples and concentrations of the dopants influence the sizes of nanoparticles in the two-phase samples (one can compare concentration changes in Figs 1 and 2).

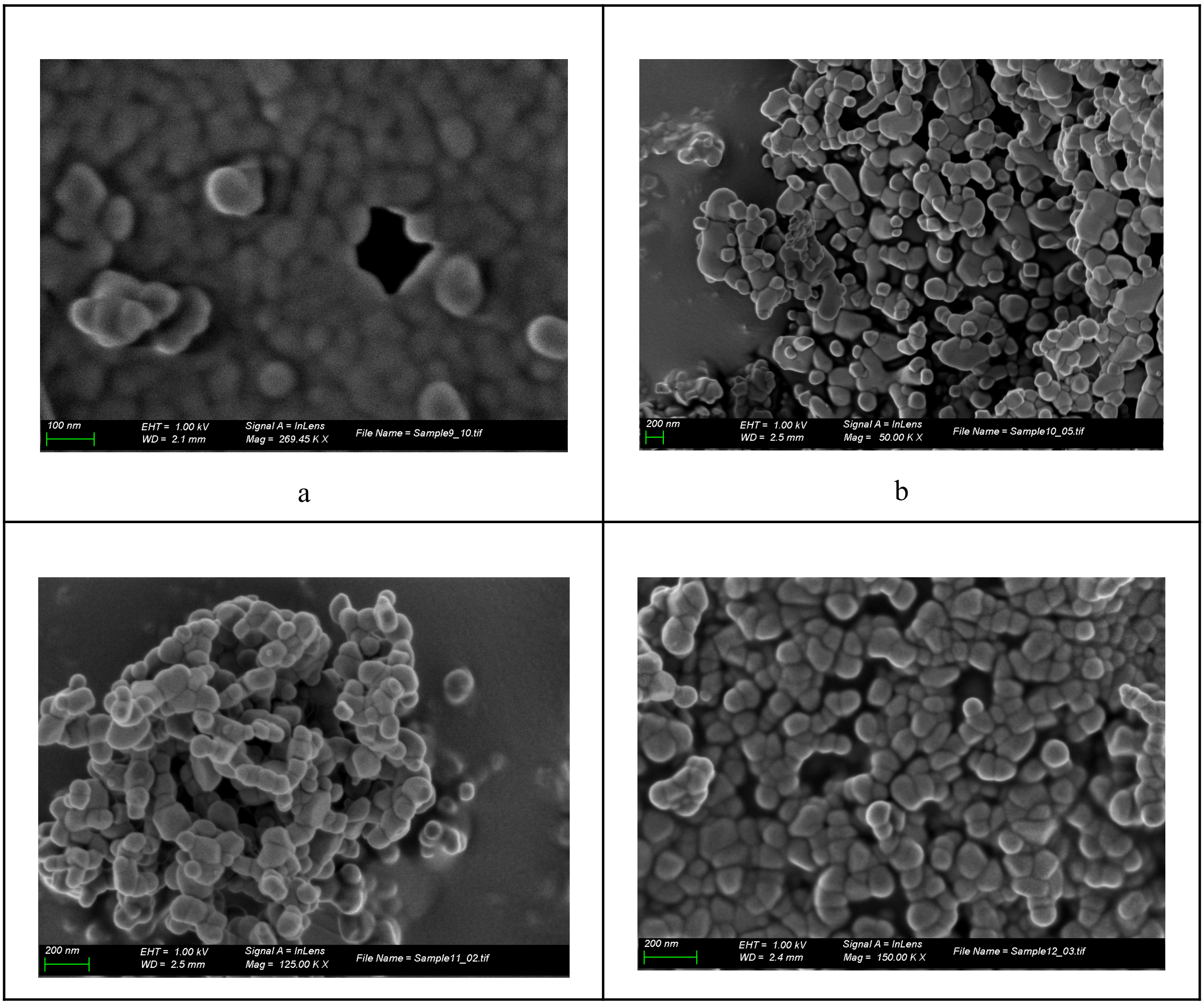

| c | d |
|---|---|

Fig. 2. SEM images of the $La_{0.9}Er_{0.025}Eu_{0.025}Ca_{0.05}VO_4$ (a), $La_{0.8}Er_{0.05}Eu_{0.05}Ca_{0.1}VO_4$ (b), $La_{0.65}Er_{0.1}Eu_{0.1}Ca_{0.15}VO_4$ (c) and $La_{0.5}Er_{0.15}Eu_{0.15}Ca_{0.2}VO_4$ nanoparticles (d).

### 3.3. Reflection spectroscopy

With the aim to study influence of $Er^{3+}$ ions on optical properties of the synthesized samples we have analyzed the spectra of diffusion reflection of the erbium containing $La_{1-x-y}Er_{x/2}Eu_{x/2}Ca_yVO_4$ samples in comparison with spectra of the un-doped ($LaVO_4$) and doped with europium ($La_{1-x}Eu_xVO_4$) and $Eu^{3+}$, $Ca^{2+}$ co-doped ($La_{1-x-y}Eu_xCa_yVO_4$) nanoparticles (Fig. 3). Diffuse reflection spectra of un-doped samples shows the main absorption band in in range 250 – 500 nm (Fig. 3, curve 5). The spectra in a diapason of minimal reflection are characterized by at least two components lying in the ranges 250 - 300 and 300 – 320 nm. The near UV and visible spectral interval of the band is characterized by sharp edge with 50 % reflection at 350 nm and weak intensity shoulder in the range 350 – 450 nm.

The spectra of europium containing samples reveal line-holes of very low intensity near 299, 318, 350, 400, and 465 nm. Spectra of the $La_{1-x-y}Er_{x/2}Eu_{x/2}Ca_yVO_4$ nanoparticles also show additional narrow spectral features peaking at 379, 523 nm and group of the lines at 653, 659 and 665 nm as well as additional wide bands in the 350 – 630 nm range. Intensity of noted narrow lines for erbium containing samples increases with increasing concentration of dopants. Spectral distributions of reflection in additional wide bands also depend on concentration of dopants and consist of two components peaking around 380 and 500 nm.

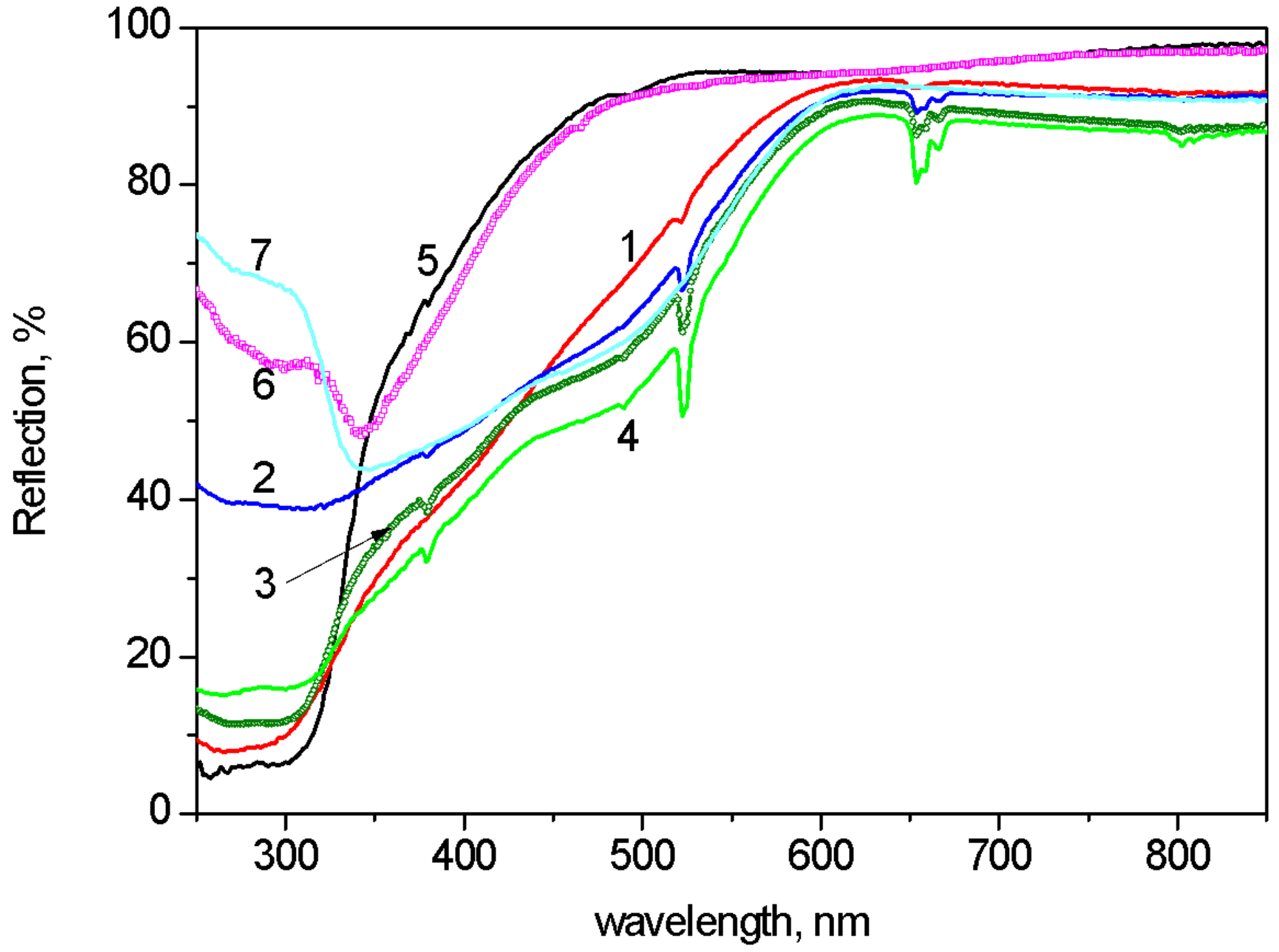

Fig. 3. Diffuse reflection spectra of the $La_{0.9}Er_{0.025}Eu_{0.025}Ca_{0.05}VO_4$ (1), $La_{0.8}Er_{0.05}Eu_{0.05}Ca_{0.1}VO_4$ (2), $La_{0.65}Er_{0.1}Eu_{0.1}Ca_{0.15}VO_4$ (3), $La_{0.5}Er_{0.15}Eu_{0.15}Ca_{0.2}VO_4$ (4), $LaVO_4$ (5), $La_{0.9}Eu_{0.1}VO_4$ (6) and $La_{0.8}Eu_{0.1}Ca_{0.1}VO_4$ (7) nanoparticles

The wide UV band observed in the reflection spectra of the un-doped $LaVO_4$ nanoparticles (Fig. 3, curve 5) is in agreement by their shape and positions with the published results [20, 47 - 49]. It is known that absorption spectra of RE orthovanadate compounds in the 250 - 350 nm range originate from electron transitions in the $VO_4^{3-}$ vanadate groups [23, 24, 30, 50 - 52]. Origins of the additional wide bands in the 350 – 630 nm range is discussed in section 4.1.

As for the narrow line holes in the reflection spectra of the $La_{1-x-y}Er_{x/2}Eu_{x/2}Ca_yVO_4$ nanoparticles, the observed narrow peaks at 379, 523 nm and group of lines at 653, 659 and 665 nm can be assigned to the $^4I_{15/2} \rightarrow {}^2G_{11/2}$ (379 nm), $^4I_{15/2} \rightarrow {}^2H_{11/2}$ (523 nm) and $^4I_{15/2} \rightarrow {}^4F_{9/2}$ (653, 659 and 665 nm) inner *f-f* electron transitions in the $Er^{3+}$ ions [40, 41, 48]. Besides, the low intensive holes at 299, 318, 350, 400, and 465 nm should be assigned to absorption transitions in the $Eu^{3+}$ ions. They are caused by the $^7F_0 \rightarrow {}^5F_2$ (299 nm), $^7F_0 \rightarrow {}^5H_4$ (318 nm), $^7F_0 \rightarrow {}^5G_2$ (350 nm), $^7F_0 \rightarrow {}^5L_6$ (400 nm) and $^7F_0 \rightarrow {}^5D_2$ (465 nm) transitions [53 - 55]. It should be also emphasized here that lines corresponded to absorption transitions in the $Eu^{3+}$ ions are much more weak compared to lines corresponded to absorption transitions in the $Er^{3+}$ ions.

### 3.4. Luminescence properties

Luminescence spectra of the $La_{1-x-y}Er_{x/2}Eu_{x/2}Ca_yVO_4$ nanoparticles consist of narrow emission lines which are observed in 400 - 720 nm range (Fig. 4,a). The lines can be divided by their positions on two diapasons lying in 400 – 560 and 570 – 720 nm spectral ranges. Emission in the 570 – 720 nm diapasons is characterized by significantly higher intensity for all the samples. In general, the observed emission lines in the 400 - 560 and 570 - 720 nm spectral ranges correspond by their positions with the well-known *f-f* electron transitions in the $Er^{3+}$ and $Eu^{3+}$ ions, respectively [26-29, 39-41]. Comparison of the spectra in Fig. 4,a reveals two main directions of emission properties dependency on dopant concentrations. There are changes of total spectra intensities, which are different for the 400 – 560 and 570 – 720 nm spectral ranges, and changes of relative intensities of some lines with concentration. It is clearly seen from Fig. 4,a that intensity of emission in the 570 – 720 nm range ($I_{(570\text{-}720)}$) is strongly depended on concentration of dopants. (Attenuation coefficients are noted in Fig. 4,a.)

Concentration dependencies of emission intensity were investigated based on the measured spectra. Relationships of emission intensities on this spectral range were estimated using spectra measured one by one at the same experimental conditions. Particularly, the same angular fraction of the emission has been used at each measurement; laser excitation beam power was controlled and laser beam spot on the sample surface was the same (~ 1 mm). The total emission values were obtained by integration of the measured emission spectra in the same spectral range. Such measurements were performed for each sample at least three times. The average intensity data are presented in Fig. 4,b. Intensity of emission was estimated as increased in more than 5 times with increasing $Er^{3+}$ and $Eu^{3+}$ions concentration from 0.025 to 0.05. The next increasing of $Er^{3+}$ and $Eu^{3+}$ ions concentration to 0.1 and 0.15 causes decrease of intensity in 5 – 10 times. Total intensity of emission in the 400 – 560 nm range, $I_{(400-560)}$, has not demonstrated so strong dependency on concentration. Intensity increases in near 2 times with $Er^{3+}$ and $Eu^{3+}$ ions concentration increasing from 0.025 to 0.05 and slightly decreases for the $La_{0.65}Er_{0.1}Eu_{0.1}Ca_{0.15}VO_4$ and $La_{0.5}Er_{0.15}Eu_{0.15}Ca_{0.2}VO_4$ samples.

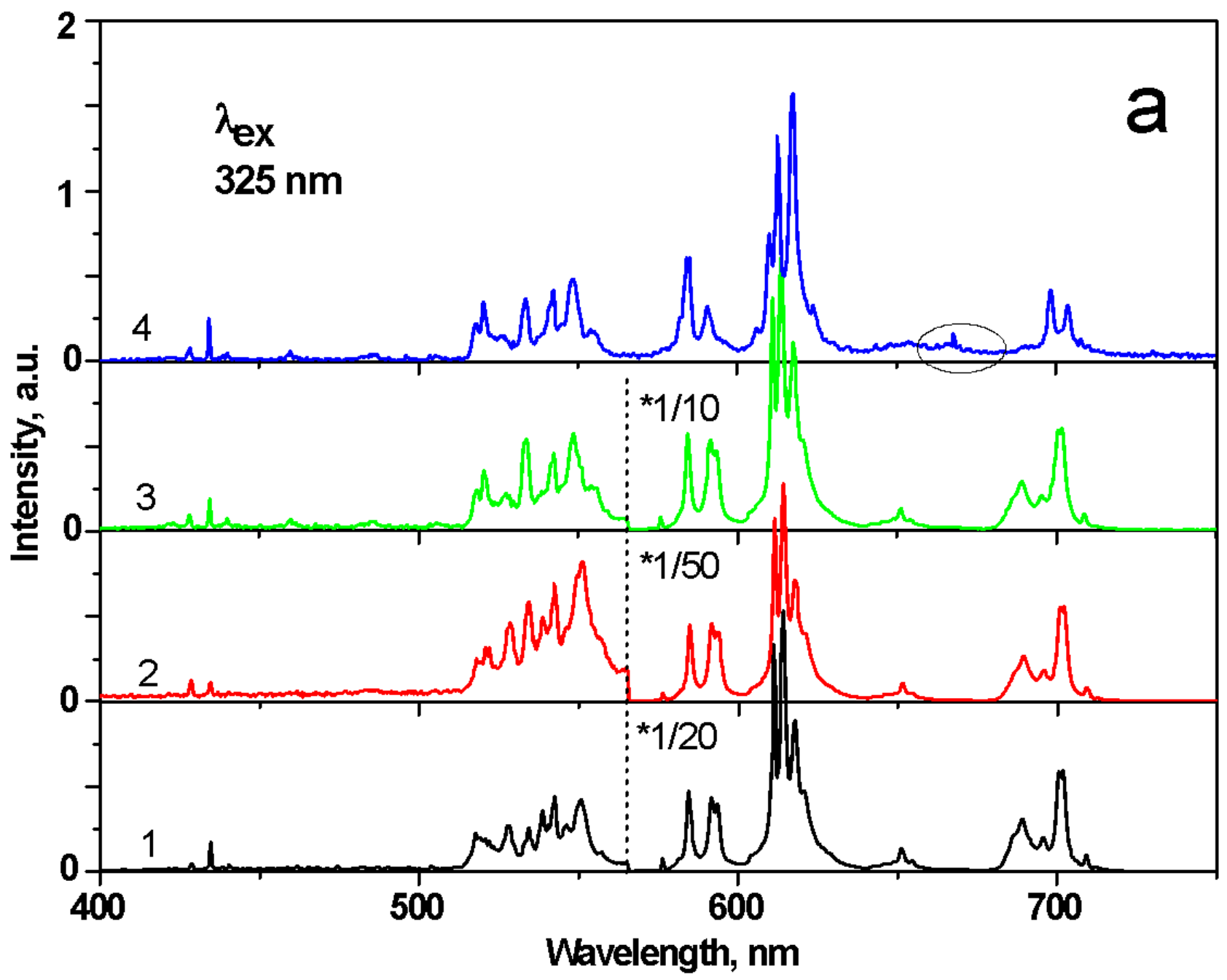

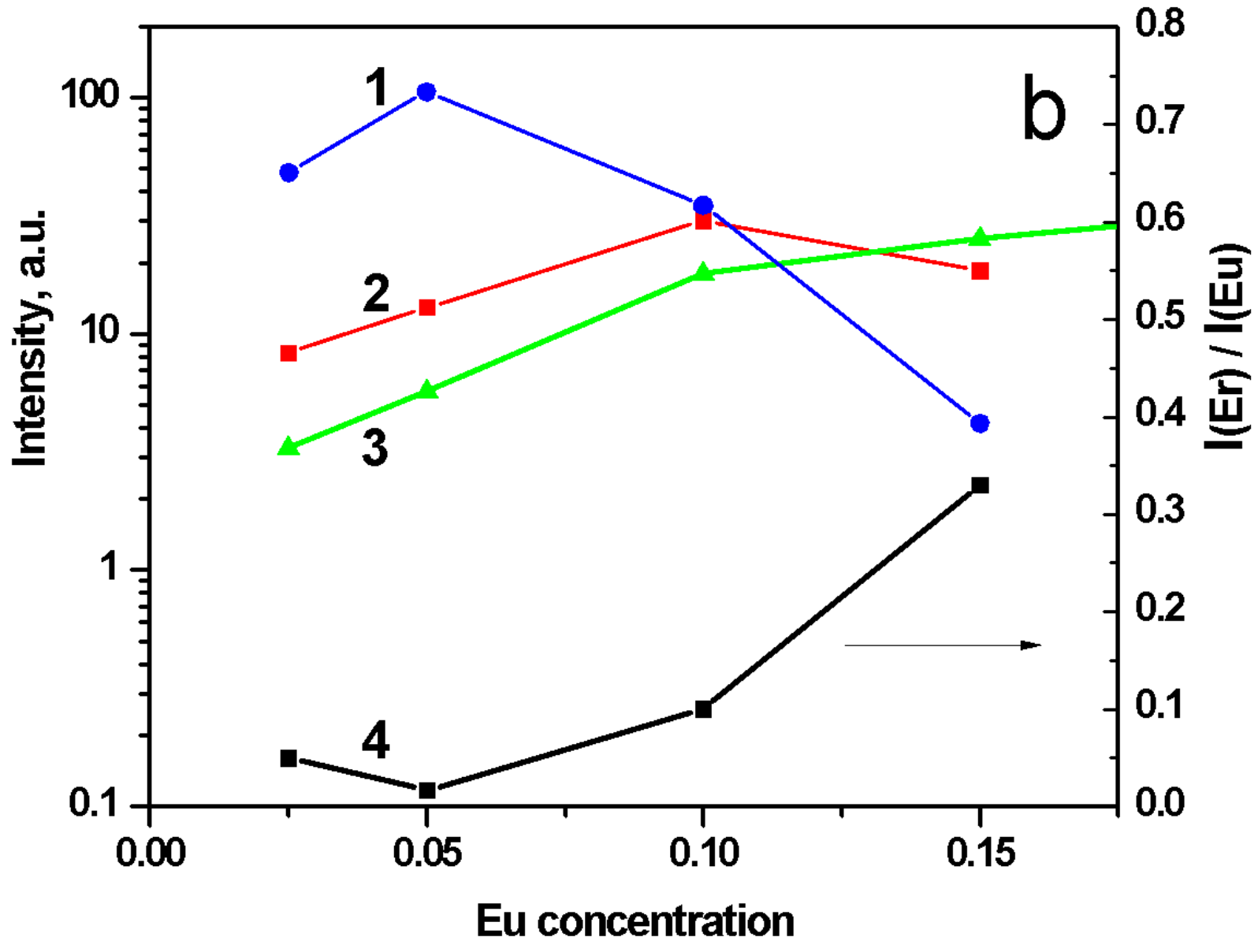

Fig. 4. a - Emission spectra of the $La_{0.9}Er_{0.025}Eu_{0.025}Ca_{0.05}VO_4$ (1), $La_{0.8}Er_{0.05}Eu_{0.05}Ca_{0.1}VO_4$ (2), $La_{0.65}Er_{0.1}Eu_{0.1}Ca_{0.15}VO_4$ (3) and $La_{0.5}Er_{0.15}Eu_{0.15}Ca_{0.2}VO_4$ (4) nanoparticles at $\lambda_{ex}$ = 325 nm

b - Dependencies of the total PL intensity over the 570-720 nm range ($I_{(570-720)}$) (1 - 3) and dependency of the $I_{(400-560)}/I_{(570-720)}$ ratio on the $Eu^{3+}$ ions concentration at $\lambda_{ex}$ = 325 nm. The samples: $La_{1-x-y}Er_{x/2}Eu_{x/2}Ca_yVO_4$ - (1, 4), $La_{1-x-y}Eu_xCa_yVO_4$ - (2), and $La_{1-x}Eu_xVO_4$ (3).

Behavior of relative intensities of various lines in the spectra with concentrations of dopants is shown in Fig. 5 in more detail. As this behavior is different for the 570 – 720 and 400 – 560 nm spectral ranges, they is considered separately in Figs 5,a and 5,b, respectively. The view of spectrum of the $La_{0.5}Er_{0.15}Eu_{0.15}Ca_{0.2}VO_4$ sample (the highest dopant concentration) in the 570 – 720 nm range is rather different compared to spectra of other three samples in Fig. 4,a. Thus, in Fig. 5,a we compare spectra of the $La_{0.5}Er_{0.15}Eu_{0.15}Ca_{0.2}VO_4$ sample and the most intensive spectra of $La_{0.8}Er_{0.05}Eu_{0.05}Ca_{0.1}VO_4$ sample. Comparing these spectra, we should mark a disappearance of the line at 576 nm and strong decrease of intensities of the lines at 584, 593, 614, 652, 686, 689, 695, 702 nm for $La_{0.5}Er_{0.15}Eu_{0.15}Ca_{0.2}VO_4$ sample if compare to the spectrum of the $La_{0.65}Er_{0.1}Eu_{0.1}Ca_{0.15}VO_4$ sample. Here and below, we will denote this set of lines as type I lines. These lines are marked by arrows in Fig. 4,a, curve 2. From the other hand, intensities of the lines at 590, 608, 612, 617, 622, 704 nm are strongly increased in the spectra of the $La_{0.5}Er_{0.15}Eu_{0.15}Ca_{0.2}VO_4$ sample. Here and below, we will denote this set of lines as type II lines. These lines are marked by asterisks in Fig. 4,a, curve 4.

Redistribution of intensities of the emission lines located within the 400 – 560 nm range has another dependency on concentration then for the lines located within the 570 – 720 nm range. The decrease of intensities of some lines is clearly seen even for the $La_{0.8}Er_{0.05}Eu_{0.05}Ca_{0.1}VO_4$ sample compared with spectra of the $La_{0.9}Er_{0.025}Eu_{0.025}Ca_{0.05}VO_4$ sample (the lowest dopant concentration). The most obvious cases of such decrease are marked by dashed boxes with arrows in Fig. 5,b. Similarly to the case described above concerning 570 – 720 nm spectral range, we will denote this set of lines, here and below, as type I lines. In the spectra of the $La_{0.65}Er_{0.1}Eu_{0.1}Ca_{0.15}VO_4$ samples, strong increase of some lines intensity was observed; especially it concerns the lines at 520, 541, 548 and 555 nm. We will denote this set of lines, here and below, as type II lines.

The dependence of emission spectra in the 570 – 720 nm range on excitation wavelength is presented in Fig. 5, c for the $La_{0.8}Er_{0.05}Eu_{0.05}Ca_{0.1}VO_4$ sample. The spectra obtained at 473 and 532 nm excitations are characterized by distribution of the lines intensity that is the same as those observed at 325 nm excitation. As for the emission spectra excited with 405 nm, it is characterized by a much lower intensity and by different distribution of lines intensity. This distribution is similar to that observed for the spectrum of $La_{0.5}Er_{0.15}Eu_{0.15}Ca_{0.2}VO_4$ sample measured at 325 nm excitation. The same lines at 590, 595, 608, 622, 705 and 708 nm are marked by asterisks in Fig. 5,a curve 4 and Fig. 5,c curve 2.

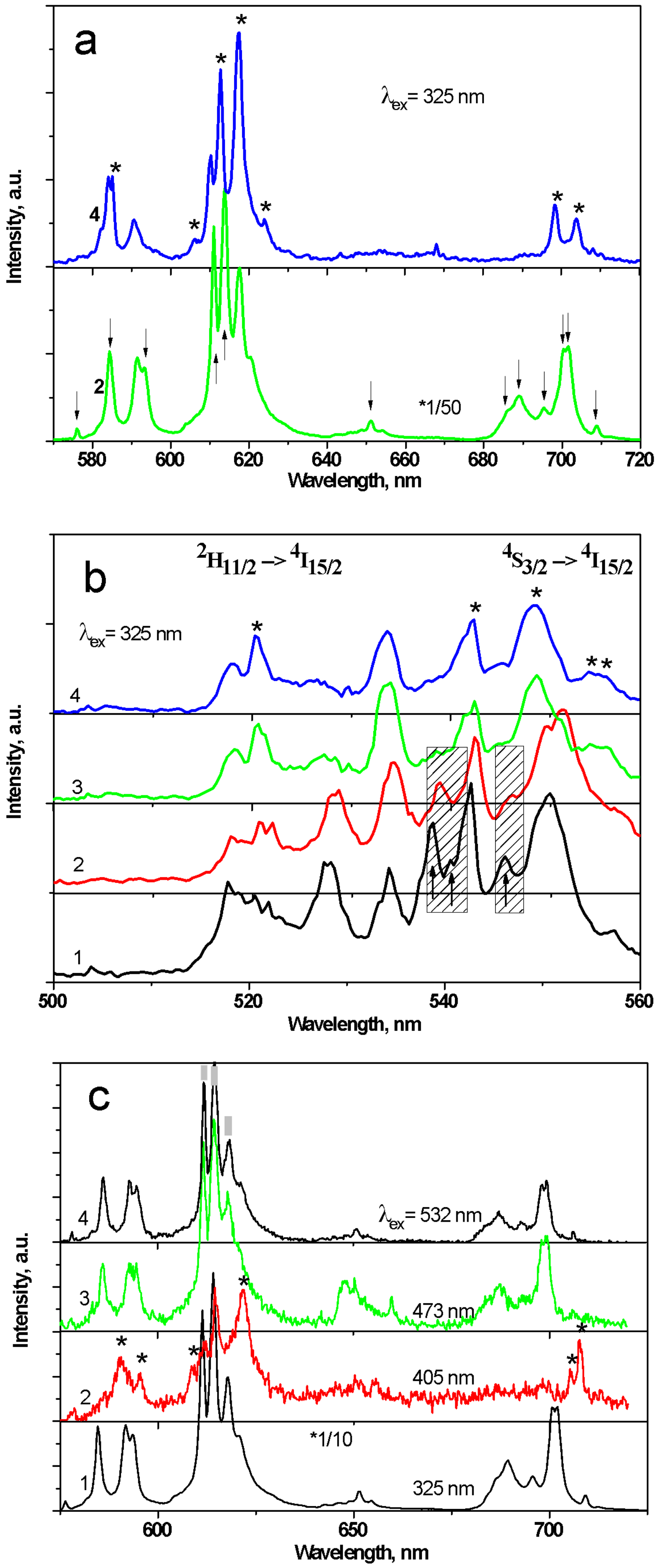

Fig. 5,a and b. Emission spectra of the $La_{0.9}Er_{0.025}Eu_{0.025}Ca_{0.05}VO_4$ (1), $La_{0.8}Er_{0.05}Eu_{0.05}Ca_{0.1}VO_4$ (2), $La_{0.65}Er_{0.1}Eu_{0.1}Ca_{0.15}VO_4$ (3) and $La_{0.5}Er_{0.15}Eu_{0.15}Ca_{0.2}VO_4$ (4) nanoparticles at $\lambda_{ex}$ = 325 nm.

Fig. 5,c. Emission spectra of the $La_{0.8}Er_{0.05}Eu_{0.05}Ca_{0.1}VO_4$ nanoparticles at $\lambda_{ex}$ = 325 (1), 405 (2), 473 (3) and 532 (4) nm excitations.

Spectra of the PL excitation were measured in the vicinity of the highest intensity lines of different behavior: 612, 614.5, and 622 nm (see gray rectangles at tops of PL peaks in the Fig. 5, c. All the excitation spectra in Fig. 6, a are very similar in their normalized view shown in Fig. 6, b. They consist of the main wide band in 250 – 360 range and spectral components above 400 nm. The main features of this part of excitation spectra coincide to details in reflection spectra, but their intensity is too weak to be discussed. It is worth noting, that relative contribution of the range above 400 nm is higher in 2 – 3 times in the spectra measured at 622 nm registration (Fig. 6, a). The position of the main band maximum for all the spectra is near 320 nm, however, we can emphasize some shift of the excitation band registered at 622 nm to the long-wavelength side compared to other spectra. The arrow-marked shoulder at about 309 nm is other feature of this spectrum (Fig. 6, c and 5, d).

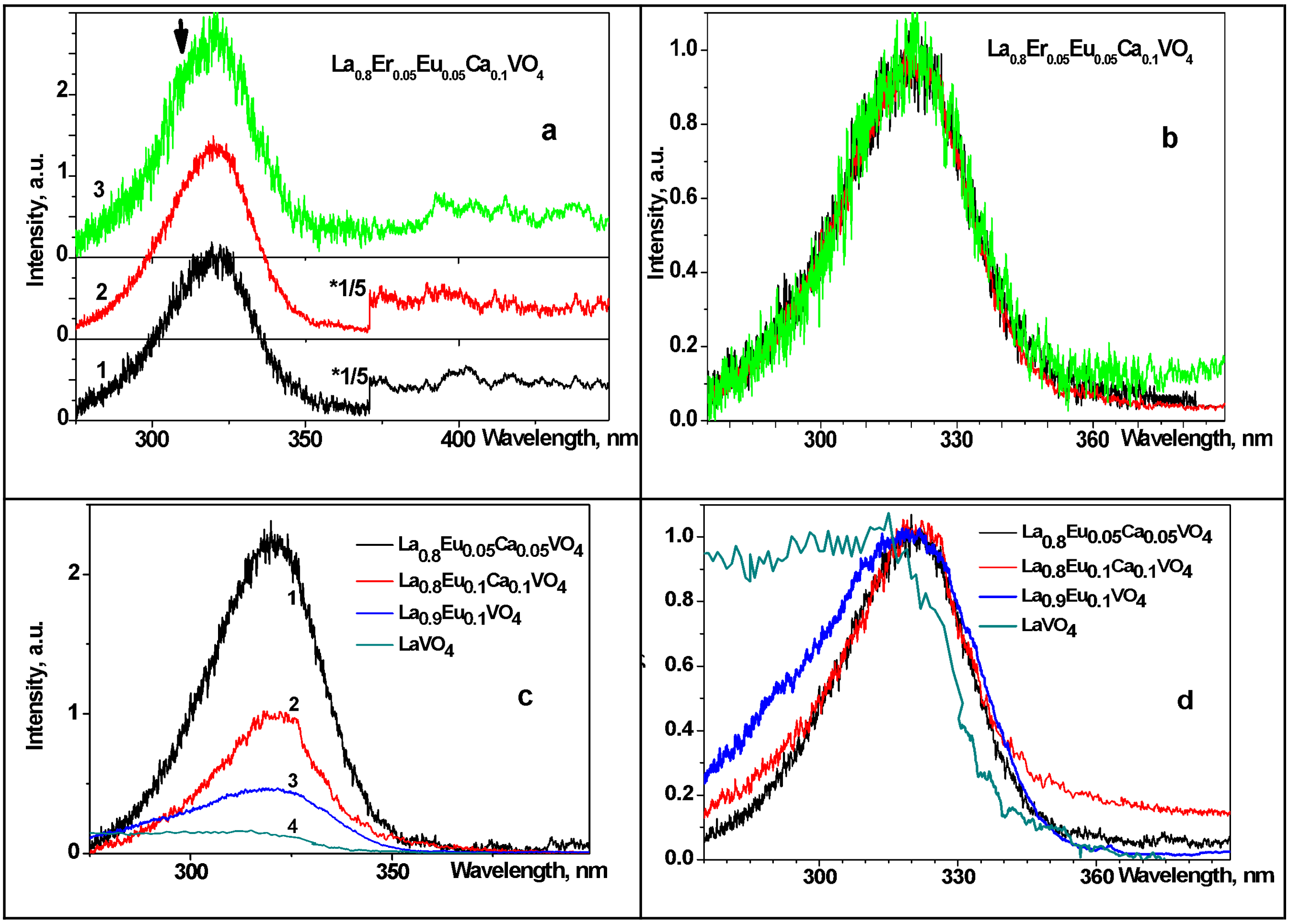

Fig. 6. a, b) Excitation spectra of the $La_{0.8}Er_{0.05}Eu_{0.05}Ca_{0.1}VO_4$ nanoparticles PL; $\lambda_{reg}$ = 612 (1), 614 (2) and 622 (3) nm (Noted normalized spectra are shown in Fig. 9b.); c, d) Excitation spectra of the

$La_{0.8}Er_{0.05}Eu_{0.05}Ca_{0.1}VO_4$ (1), $La_{0.8}Eu_{0.1}Ca_{0.1}VO_{4,}$ (2), $La_{0.9}Eu_{0.1}VO_4$ and $LaVO_4$ nanoparticles (4) at $\lambda_{reg}$ = 614 (1 - 3) and 550 (4) nm (Noted normalized spectra are shown in Fig. 9d.)

## 4. **Discussion**

### 4.1. Effect of dopants on electronic states and crystal lattice

The co-doping affects both the structure, phase composition and the optical properties of the $LaVO_4$:Eu. The doping effect depends on the concentration of dopants. In particular, the XRD results give possibility to make assumption about presence in the nanoparticles with the low dopant concentrations ($La_{0.8}Er_{0.05}Eu_{0.05}Ca_{0.1}VO_4$ sample) of traces of tetragonal phase. Detailed analysis of reflection and excitation spectra supported this assumption.

Arrangement of the $Eu^{3+}$ (effective ionic radius is 1.12 Å for nine-fold coordination) in the $La^{3+}$ sites (effective ionic radius is 1.216 Å) increases the crystal field strength in the environment of the europium ion, and as a sequence its oxygen environment and the nearest $VO_4^{3-}$ groups are deformed, as compared to the un-doped $LaVO_4$. As a result, the local states occur in the band gap near bottom of the conduction band and near top of the valence band. Thus, absorption edge is shifted to the red side and absorption increases in the 350 - 500 nm range (Fig. 3).

Effect of erbium ions on crystal and energy structure of the $LaVO_4$ matrix is similar, but more distinctive as compared to one for the $Eu^{3+}$ ions, because the radius of the $Er^{3+}$ ions is lower (1.063 Å for nine-fold coordination) [56].

The ionic radius of the $Ca^{2+}$ (1.18 Å for nine-fold coordination) is closer to size of lanthanum ion compared to $Er^{3+}$ and $Eu^{3+}$ ions. However, in addition to the change in lattice field potential at the $La^{3+} \rightarrow Ca^{2+}$ substitution, necessity of $[Ca^{2+}/La^{3+}]^-$ negative effective charge compensation occurs in this case. The compensation can be achieved by creation of $[V_O]^{2+}$ oxygen vacancies. These processes promote additional deformation of the $LaVO_4$ monoclinic crystal lattice and reduction of vanadium ions [57].

According to band structure calculations, the monoclinic $LaVO_4$ is an indirect band gap material, where similarly to other oxide compounds the top of the valence band is formed by the O 2*p* states, while the V 3*d* states dominate the bottom of the conduction bands. Moreover, a hybridization of the O 2*p* and V 3*d* states occurs both in valence and conduction bands. Thus, absorption transitions in $LaVO_4$ are accompanied by energy transfer from oxygen ligands to the vanadium ions: $O^{2-} \rightarrow V^{5+}$ [24, 31]. ($C_1$ and $D^2{}_d$ for monoclinic and tetragonal phase, respectively). The $VO_4^{3-}$ group symmetry in the $LaVO_4$ is lower than for initial free $VO_4^{3-}$ anion ($C_1$ and $D_{2d}$ for monoclinic and tetragonal phase, respectively). Due to the decrease of symmetry, the excited states of the $VO_4^{3-}$ group split, and therefore the mentioned absorption band is complex. According to experimental and calculated estimations, the band gap of monoclinic $LaVO_4$ (direct band to band transitions) is estimated by value of 3.5 eV (354 nm) [24, 31, 49], that approximately corresponds to the observed sharp edge of the reflection band (Fig. 3). Therefore, long wave length shoulder (350 – 630 nm range) observed in the reflection spectra of some doped samples should be assigned to transitions on the levels lying in the forbidden band. Taking into account that reflection bands in the 350 – 630 nm range were observed only for the Ca-doped samples, they must be connected with Ca-induced defects. The most probably that such defect states as $V^{4+}$ and $V^{3+}$ ions, which are formed under synthesis [58, 59] and locate at the $V^{5+}$ sites of the regular orthovanadate lattice, are responsible for the additional bands around 380 and 500 nm, respectively (Fig. 3, curves 1 - 4). We understand that despite the data presented in [58, 59], it would be better to confirm them with our own experimental data, e.g. by XPS investigation of the studied materials. That is why, the XPS experiments are in our plans for the next study.

Effects of the $Er^{3+}$ and $Ca^{2+}$ ions on electronic states, on defects of crystal structure and on formation of additional absorption bands can be observed also in excitation spectra of the $La_{1-x-y}Er_{x/2}Eu_{x/2}Ca_yVO_4$ nanocrystals. Exhibition of the connected with the $V^{4+}$ defects additional absorption band around 380 nm in the excitation spectra of the investigated nanoparticles (Fig. 6,a) reveal that these defects take part in the luminescence process. Also, we should note that excitation of the RE ions luminescence in orthovanadates is effective just in the matrix absorption bands. It is known that transfer of excitation energy from $VO_4^{3-}$ groups to the RE ions is a mechanism of such excitation [29, 30, 38, 39], and the data shown above at Figs. 3 and 6 confirm that. We can see that excitation spectra of the un-doped $LaVO_4$ (Fig. 6, curve 1) correspond with reflection band in the 250 – 350 nm range (Fig. 3, curve 5). At the contrary to excitation of own emission, the $Eu^{3+}$ emission is effectively excited mainly in the long wave length range of the $LaVO_4$ absorption (Fig. 6, curve 2). Thus, that band should be interpreted as caused by creation and radiation decay of localized excitons [24, 50]. This evidences about essential non-radiation losses during the process of $VO_4^{3-} \rightarrow VO_4^{3-}{}_{\dots} \rightarrow VO_4^{3-}{}_{\dots} \rightarrow Eu^{3+}$ energy migration along the lattice. The noted losses are more noticeable for the $La_{1-x}Eu_xCa_yVO_4$ and $La_{1-x-y}Er_{x/2}Eu_{x/2}Ca_yVO_4$ compositions (Fig. 6,d, curves 3, 4). The observed changes confirm the above made assumption about increased role of the lattice deformations and defects in determination of optical, in particular luminescent, properties of the $La_{1-x-y}Er_{x/2}Eu_{x/2}Ca_yVO_4$ nanocrystals compared to the $La_{1-x}Eu_xVO_4$ ones.

### 4.2. Influence of phase crystal structure on luminescent properties

Effects of the Er, Ca co-doping on efficiency and intensity of the $La_{1-x-y}Er_{x/2}Eu_{x/2}Ca_yVO_4$ nanoparticles are determined by several factors. Above we have considered the excitation energy losses on the defect induced traps. From the other hand, significant increasing of emission intensity in the samples with low content of co-dopants has been described in the section 3. Moreover, intensity of excitation of the $Eu^{3+}$ emission essentially increases in a sequence of the $La_{1-x}Eu_xVO_4 \rightarrow La_{1-x}Eu_xCa_yVO_4 \rightarrow La_{1-x-y}Er_{x/2}Eu_{x/2}Ca_yVO_4$ compositions (Fig. 6,c). Such enhancement can be caused as by changes of crystal field and symmetry of the $Eu^{3+}$ local surrounding, by composition and structure of the $Eu^{3+}$ emission centers, as well as by local symmetry in the vicinity of the $Eu^{3+}$ ions, and in volume phase transformation of the $LaVO_4$ crystal lattice.

Developed by us the model of formation of the complex emission centers in the $La_{1-x}Eu_xCa_yVO_4$ and $Eu_{1-x}Ca_xVO_4$ nanoparticles doped with calcium ions [43, 44] takes into account such local transformation. In accordance with that model, such centers are formed by the $Eu^{3+}$ ions and neighbor $Ca^{2+}$ ions, and oxygen vacancies. Intensity of such centers luminescence is higher, compared to intensity of emission of the simple $Eu^{3+}$ centers [60], due to deformation of local crystal field around the neighbor $O^{2-}$ ions [29]. Distinctive dependence of emission intensity of the $Eu^{3+}$-doped $LaVO_4$ nanocrystals on geometry and crystal field strength has been also reported previously, e.g., in a work devoted to study of the alkali-earth ions and anisotropy of the shapes of nanocrystals effects on characteristics of their emission [20, 39].

Another important aspect of the noted transformation is change of distribution of atoms in the nearest surrounding of the $RE^{3+}$ and $V^{5+}$ ions. It is known that in the monoclinic $LaVO_4$ crystal lattice among the 9 La–O bond lengths and bond populations (2.465, 2.472, 2.480, 2.484, 2.505, 2.536, 2.629, 2.671, 2.897 Å and 0.22, 0.21, 0.21, 0.18, 0.16, 0.15, 0.11, 0.10, and 0.02 e, respectively [31]). The 2.897 Å bond length and 0.02 e population are essentially different from the other 8 pairs of parameters. When RE dopants substitute for the $La^{3+}$, the oxygen atoms in their environment should be shifted toward $Eu^{3+}$ or $Er^{3+}$ ions. Obviously, different oxygen atoms undergo different shifts, and the most distant one will "break away" from the other eight. Therefore, the geometry structure similar to the case of the tetragonal $LaVO_4$ can arise in the $Eu^{3+}$ or $Er^{3+}$ surrounding (only two values of the La–O bond lengths: 2.441 and 2.523 Å; only two values of bond population: 0.21 and 0.09 [31]). In such a case, an oxygen vacancy can probably be formed in the volume of crystal. It is known, the deep acceptor levels in forbidden band of $REVO_4$ are related with those oxygen vacancies. Thus, their formation can cause the observed shift of the excitation and absorption spectra to the long wave length side (see Fig. 2, b and Fig. 6).

In a typical for the t-$LaVO_4$ phase surrounding of the $Ln^{3+}$ ions, the PL efficiency may increase due to more effective $VO_4^{3-} \rightarrow Eu^{3+}$ local transfer of excitation energy. Indeed, in line with the discussion by G. Blasse [29], and with the following it application in a number of subsequent publications [23, 26, 30, 41], increasing of the bond angle from 90° and higher leads to more efficient transfer of the excitation energy from the $VO_4^{3-}$ group to the RE ion. This feature is caused by the characteristic set of molecular orbitals of the $VO_4^{3-}$ group. The V and O atoms in the $VO_4^{3-}$ molecular anion are connected by four types of covalence interactions: there are σ bonding, π bonding, π* antibonding and σ* antibonding. Exchange interaction is higher and as a sequence an excitation energy transfer is the most effective for a case of σ bonding (angle is 180°) compared with π bonding (angle is 90°). For the $Eu^{3+}/La^{3+}$ site in the m-$LaVO_4$ phase, only one V-O-$Eu^{3+}$ angle for 9 bonds is much higher than 90° (153°), whereas al the other angles are essentially smaller, they are from 89° to 137° [23]. For the $Eu^{3+}/La^{3+}$ site in the t-$LaVO_4$ phase, four from 8 bonds have the V-O-$Eu^{3+}$ angle equal 153° (four other angles are equal 98°). That leads to essential increase of excitation energy transfer from the $VO_4^{3-}$ groups to the $Eu^{3+}$ or $Er^{3+}$ ions when their surrounding is transforming from monoclinic to tetragonal-like.

Reliable confirmation of the lattice transformation from m-$LaVO_4$ to t-$LaVO_4$ even at low RE impurity concentrations is provided by the analysis of luminescence spectra.

Above, we pointed out, that the narrow lines in the emission spectra of the $La_{1-x-y}Er_{x/2}Eu_{x/2}Ca_yVO_4$ samples are caused by well-known inner *f-f* electron transitions in the $Er^{3+}$ and $Eu^{3+}$ ions [39 – 41, 60 – 62]. These are $^5D_0 \rightarrow {}^7F_J$ (J = 0, 1, 2, 3, 4) radiation transitions in the $Eu^{3+}$ ions (570 – 720 nm range) [39, 63] and $^4F_{7/2}$, $^2H_{11/2}$, $^4S_{3/2} \rightarrow {}^4I_{15/2}$ radiation transitions in the $Er^{3+}$ ions (420 – 560 nm range) [44, 62]. The group of the very weak intensity lines (660 – 690 nm range) with the main peak at 668 nm, that was clearly observed only for the sample with higher concentration of $Er^{3+}$ ions (indicated in Fig. 4, a), should be also assigned to the $^4F_{9/2} \rightarrow {}^4I_{15/2}$ transition in the $Er^{3+}$ ions.

The lines of the $Eu^{3+}$ emission show considerably higher intensity at 325 nm excitation than the lines of the $Er^{3+}$ ions, that especially concerns $La_{0.8}Er_{0.05}Eu_{0.05}Ca_{0.1}VO_4$ sample, where intensity of emission of the $Eu^{3+}$ ions is in 60 times higher than intensity of the $Er^{3+}$ ions emission (Fig. 4, a, curve 2).

Observed redistribution of the lines in the emission spectra with concentration is obviously caused by existence of two crystal phases in the samples under study. Similar effects have been described previously, not only for $LaVO_4$ nanocrystals doped with $Eu^{3+}$ [6], but also for $LaVO_4$ nanocrystals doped with $Sm^{3+}$ ions [46]. As for the physical mechanisms of these effects, they should be associated with difference in symmetry of the crystal surrounding of the RE ions in the monoclinic and tetragonal phases of the $La_{1-x-y}Er_{x/2}Eu_{x/2}Ca_yVO_4$ crystal lattice.

It is quite difficult to undoubtedly distinguish all the PL lines, as belonging to the emission of the RE ions in the different crystal phases because of the significant overlapping of the lines. However, some of them can be clearly assigned to different types of the $Eu^{3+}$ PL centers: type-I or type-II which emit the linear spectra of the type I and type II, respectively.

Observation of 578 nm line corresponding to $^5D_0 \rightarrow {}^7F_0$ transition (see Fig. 4, a) for all the samples except the $La_{0.5}Er_{0.15}Eu_{0.15}Ca_{0.2}VO_4$ confirms that type-I centers are characterized by low symmetry of the RE ions surrounding These centers should be assigned to the $Eu^{3+}$ ions arranged in the nanoparticles with monoclinic crystal lattice.

Emission spectra of the $La_{0.5}Er_{0.15}Eu_{0.15}Ca_{0.2}VO_4$ sample, which is of single tetragonal phase, contain only lines of the type-II. This fact confirms that type-II centers are characterized by higher symmetry of the RE ions surrounding and could be assigned to the $Eu^{3+}$ ions arranged in the nanoparticles with tetragonal crystal lattice. We clearly see also that the type-II centers are in the $La_{0.65}Er_{0.1}Eu_{0.1}Ca_{0.15}VO_4$ samples (Fig. 4, a).

The lines of the $Eu^{3+}$ ions emission in t-$LaVO_4$ nanoparticles are also in the spectra of the sample of lower dopants content - $La_{0.8}Er_{0.05}Eu_{0.05}Ca_{0.1}VO_4$. The weak but clear emission of the type-II $Eu^{3+}$ centers can be registered for this sample at 405 nm laser excitation (see Fig. 5, c, curve 2).

The t-$LaVO_4$ phase was also found for the $La_{0.9}Er_{0.025}Eu_{0.025}Ca_{0.05}VO_4$ sample, where concentrations of the $Er^{3+}$ and $Eu^{3+}$ are the lowest. An analysis of the $Er^{3+}$ emission spectra permitted us to do this conclusion (Fig. 5, b). In fact, similarly to the case of the $Eu^{3+}$ emission it was also possible to divide the $Er^{3+}$ emission lines into two sets as belonging to the luminescence centers of type-I and type-II. However, the lines of the $Er^{3+}$ II type centers (t-$LaVO_4$ phase) were measured for the $La_{0.9}Er_{0.025}Eu_{0.025}Ca_{0.05}VO_4$ sample, while $Eu^{3+}$ type-II centers (t-$LaVO_4$ phase) could be distinguished only for the sample with higher concentration of dopants, that is $La_{0.9}Er_{0.05}Eu_{0.05}Ca_{0.1}VO_4$ sample.

It is well known that transformation of monoclinic $LaVO_4$ phase into tetragonal with RE-doping is connected with replacement of large-radius $La^{3+}$ ions by RE ions of smaller radii [24, 28, 34, 37]. As ionic radius of the $Er^{3+}$ ions is smaller that ionic radius of the $Eu^{3+}$ ions, then transformation into the tetragonal phase and stabilization of the last one have to take place for the $Er^{3+}$ dopants with lower concentrations compared to the $Eu^{3+}$ dopants. It is obvious, a cumulative effect occurs under co-doping with $Eu^{3+}$ and $Er^{3+}$ ions. Thus, t-$LaVO_4$ begins to form at their concentration near 2.5 mol. %. This is a fairly low level of doping, which already causes the formation of the tetragonal phase, compared with the known ones. As a rule, specific $Eu^{3+}$ concentration, when the $LaVO_4$ tetragonal phase is fixed on the background of monoclinic phase, reaches values near 10 – 20 mol. % [21, 23, 35]. Only in colloidal solutions, or if special reagents similar to chelating ligands are used at the synthesis, the controlled creation of t-$LaVO_4$ is carried out at the wide range of the $Eu^{3+}$ content.

**4.3. Mechanisms of luminescence intensity dependences on dopants concentration**

The cumulative effect of factors influencing the luminescence intensity determines also the concentration dependence of emission intensity of the $Eu^{3+}$ centers. In fact, increase of the $Eu^{3+}$ concentration in 2 times (from 2.5 up to 5 mol. %) leads increasing intensity of the $La_{1-x}Eu_xVO_4$ and $La_{1-x-y}Eu_xCa_yVO_4$ nanoparticles in ~ 1.6 times, whereas for the samples co-doped with the $Er^{3+}$ and $Ca^{2+}$ ($La_{1-x-y}Er_{x/2}Eu_{x/2}Ca_yVO_4$) intensity of the $Eu^{3+}$ emission increases in 2.2 times (Fig. 4,b, curve 1). This difference is a result just of the cumulative effect of the factors both the crystal field enhancing in complex PL centers and of transformation of crystal phases (m-$LaVO_4$ → t-$LaVO_4$) in the $La_{1-x-y}Er_{x/2}Eu_{x/2}Ca_yVO_4$ nanoparticles. As shown above in Section 4.2, symmetry of crystal field surrounding strongly affects on the probabilities of emission transitions in the $RE^{3+}$ ions. The probabilities are increased for the RE ions in monoclinic m-$LaVO_4$ crystal lattice, especially when $Ca^{2+}$ ions were added. As results, crystal surrounding is characterized by lower symmetry that usually leads to increase of intensities of the corresponded emission transitions. From the other hand, the rate of energy transfer from host to activator ions is also effects on luminescence intensity and it is higher for the tetragonal t-$LaVO_4$ crystal lattice. We consider that just combination of two phases have allowed us to obtain such considerable increase of luminescence intensity: tetragonal phase provides high rate of excitation energy transfer to the activator RE ions and monoclinic phase provides high rate of emission transitions. Indeed, the next increase of the $Eu^{3+}$ concentration in 2 times (from 5 up to 10 mol. %) is followed by increase of emission intensity of the $La_{1-x}Eu_xVO_4$ and $La_{1-x-y}Eu_xCa_yVO_4$ nanoparticles in ~ 3.0 times that undoubtedly should be caused by increase of the t-$LaVO_4$ phase content which determines higher energy transfer efficiency of the both space ($VO_4^{3-}{}_{\dots}$ → $VO_4^{3-}$) and local ($VO_4^{3-}$ → $Eu^{3+}$) types compared to the monoclinic phase.

A concentration quenching of the $Eu^{3+}$ emission starts already at mentioned above concentration range for the $La_{1-x-y}Er_{x/2}Eu_{x/2}Ca_yVO_4$ samples (Fig. 4,b, curve 1). Prevail of quenching over intensity rising is a result both of formation of a lot of traps ($Ca^{2+}$ ions concentration noticeably increased) and increase of tetragonal phase content. Noted, that probability of excitation energy migration is higher for t-$LaVO_4$ phase. So, the probability of nonradiative energy loss at the migration increases in a result of common effect of the both mentioned factors.

This effect if also observed for the $La_{1-x-y}Eu_xCa_yVO_4$ samples, but in the next concentrations range from 10 up to 15 mol. % (Fig. 4,b, curve 2). Only for the samples of the $La_{1-x}Eu_xVO_4$ series quenching is not observed in the 2.5 – 20 mol. % concentration range, but rise of intensity is slowed to 1.4 and 1.3 for the 10 – 15 and 15 – 20 mol. % ranges, respectively (Fig. 4,b, curve 2). The maximum intensity was found for the $La_{0.8}Er_{0.05}Eu_{0.05}Ca_{0.1}VO_4$ composition. The quoted value is in 8 ± 2 times higher than that of the $La_{0.9}Eu_{0.05}Ca_{0.05}VO_4$ compound and in 19 ± 2 times higher than the luminescence intensity of the $La_{0.95}Eu_{0.05}VO_4$ compound, where the $Eu^{3+}$ ions concentration is the same (Fig. 4,b). This notes that number of defects in these samples increases slower with concentration comparably to the samples additionally doped with erbium and calcium ions.

One more aspect of excitation energy transformation is reflected in dependence of the ratio of the $Er^{3+}$ and $Eu^{3+}$ ions PL intensity ($I_{Er}/I_{Eu}$) on dopant concentrations (Fig. 4,b, curve 4). This non-monotonic dependence may be caused by energy transfer from the $Er^{3+}$ to $Eu^{3+}$ ions in the $La_{1-x-y}Er_{x/2}Eu_{x/2}Ca_yVO_4$ nanocrystals, we suppose.

It is known that several lines of excitation of the $Eu^{3+}$ emission are located in the ranges (395 – 430, 505 – 535, and 550 – 560 nm) of the $^2H_{9/2} \rightarrow {}^4I_{15/2}$, $^2H_{11/2} \rightarrow {}^4I_{15/2}$, and $^4S_{3/2} \rightarrow {}^4I_{15/2}$, emission transitions in the $Er^{3+}$ ions [40, 41, 48]. Therefore, energy transfer can take place in the sequences $VO_4^{3-}{}_{\dots} \rightarrow VO_4^{3-} \rightarrow Er^{3+}{}_{\dots} \rightarrow Eu^{3+}$. The scheme of the possible electron radiation and radiation-less transitions in the $La_{1-x-y}Er_{x/2}Eu_{x/2}Ca_yVO_4$ system is presented in Fig. 7.

Excitation energy transfer from matrix to the $Er^{3+}$ ions should be influenced by the same factors which have been above discussed about $Eu^{3+}$excitation and luminescence. So, the $Er^{3+} \rightarrow Eu^{3+}$ excitation energy transfer can enhance the intensity of the $Eu^{3+}$ ions and lower intensity of the $Er^{3+}$ ions emission. Thus, a decrease of the $Er^{3+}$ PL relative intensity on the 2.5 to 5 mol. % step of the $Eu^{3+}$ concentration can be caused by intensification of the $Er^{3+} \rightarrow Eu^{3+}$ energy transfer. A lot of interceptions of excitation energy by traps decrease the role of such energy transfer in the $La_{1-x-y}Er_{x/2}Eu_{x/2}Ca_yVO_4$ system at higher concentrations of dopants. As a result, the $I_{Er}/I_{Eu}$ intensity ratio increases.

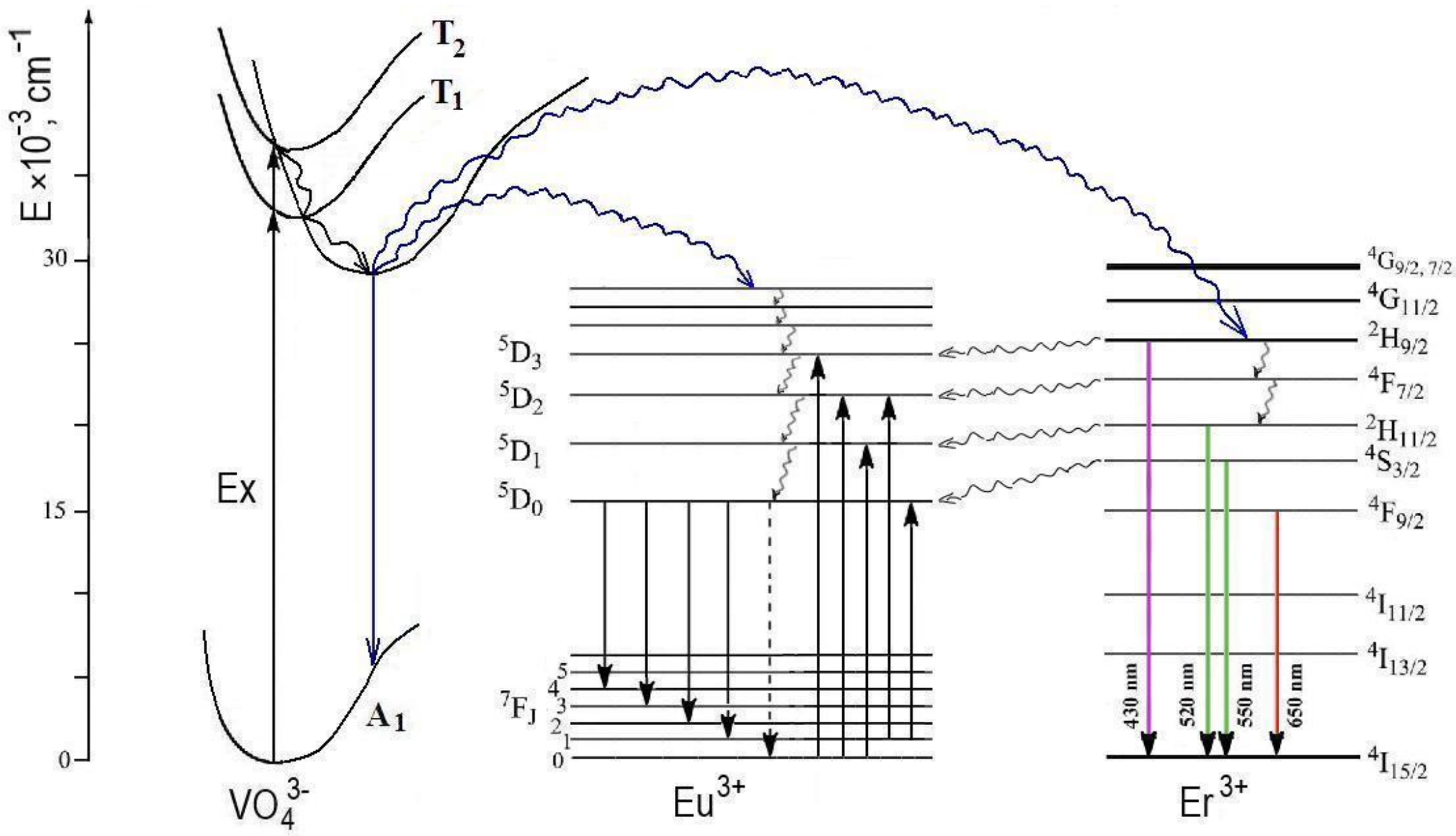

Fig. 7. Scheme of electron transitions in the $La_{1-x-y}Er_{x/2}Eu_{x/2}Ca_yVO_4$ system

## 5. Conclusions

The $LaVO_4$:Eu crystalline nanoparticles were synthesized and investigated in this work with the goal to find compositions with enhanced intensity of the $Eu^{3+}$ ions luminescence and to study the mechanisms of concentration effects of various impurities on their luminescence properties. Impact of the $Er^{3+}$ and $Ca^{2+}$ ions on the luminescent properties of the $LaVO_4$ nanocrystals doped with $Eu^{3+}$ ions was studied on the set of the $La_{1-x-y}Er_{x/2}Eu_{x/2}Ca_yVO_4$ ($0 \leq x \leq 0.3$, $0 \leq y \leq 0.2$) nanocrystals. Crystal phase composition, morphology of the samples and light reflection spectra were studied and relation of these properties with luminescence characteristics were discussed.

The $Er^{3+}$ and $Ca^{2+}$ dopants promote crystal phase transformation of the doped $LaVO_4$ nanocrystals. An analysis of the XRD and luminescence data showed that the $La_{1-x-y}Er_{x/2}Eu_{x/2}Ca_yVO_4$ samples with the lower dopant concentration (x = 0.05 and 0.1) are crystallized mainly in a monoclinic monazite-type structure (m-$LaVO_4$) with secontary phase of the tetragonal zircon-type structure (t-$LaVO_4$). The increase of dopant concentrations leads to increase of the tetragonal phase content, and the $La_{0.5}Er_{0.15}Eu_{0.15}Ca_{0.2}VO_4$ samples are single phase of t-$LaVO_4$ structure.

Two types of photoluminescence centers located in monoclinic and tetragonal crystal phases were recognized owing to narrow lines of their emission spectra caused by *f-f* radiation transitions in the $Eu^{3+}$ and $Er^{3+}$ ions.

The cumulative effect of dopants on the structure and defects of the crystal lattice, neighborhood environment of $Eu^{3+}$ ions and the efficiency of excitation energy transfer both along the vanadate matrix and from $VO_4^{3-}$ molecular groups to $Eu^{3+}$ ions determines the dependence of $Eu^{3+}$ ions luminescence intensity on $Er^{3+}$ and $Ca^{2+}$concentrations. The maximum intensity was found for the $La_{0.8}Er_{0.05}Eu_{0.05}Ca_{0.1}VO_4$ composition. The quoted value is in 8 ± 2 times higher than that of the $La_{0.9}Eu_{0.05}Ca_{0.05}VO_4$ compound and in 19 ± 2 times higher than the luminescence intensity of the $La_{0.95}Eu_{0.05}VO_4$ compound, where the $Eu^{3+}$ ions concentration is the same.

The $Er^{3+}$ and $Ca^{2+}$ ions co-doping is a promising way to increase the luminescence efficiency of the $Eu^{3+}$ ions in $LaVO_4$ nanocrystals. At the same time, additional research is still needed to elucidate the characteristics of excitation energy transfer in systems of high complexity, such as the $La_{1-x-y}Er_{x/2}Eu_{x/2}Ca_yVO_4$ nanocrystals. These may be studies of the luminescence decay and effects of temperature on luminescence characteristics.

**Acknowledgements**

This work has received funding from Ministry of Education and Science of Ukraine the Polish National Agency for Academic Exchange (Bilateral Grant Program) and from the Horizon Europe research and innovation program under grant agreement No 654360 having benefited from the access provided by Institute of Electronic Structure & Laser (IESL) of Foundation for Research & Technology Hellas (FORTH) in Heraklion, Crete, Greece and Division of Microscopy, Universitat Autonoma de Barcelona, Spain within the framework of the NFFA-Europe Transnational Access Activity.

**Conflict of interest statement**

Authors declare that there are no conflicts of interest.

**References**

[1] G. Panayiotakis, D. Cavouras, I. Kandarakis, C. Nomicos, A study of X-ray luminescence and spectral compatibility of europium-activated yttrium-vanadate ($YVO_4$:Eu) screens for medical imaging applications, Appl. Phys. A., 62 (1996) 483-486. https://doi.org/10.1007/BF01567121.

[2] F.C. Palilla, A.K. Levine, M. Rinkevics, Rare Earth Activated Phosphors Based on Yttrium Orthovanadate and Related Compounds, J. Electrochem. Soc. 112 (1965) 776-779. https://doi.org/10.1149/1.2423693.

[3] A.H. Krumpel, E. van der Kolk, E. Cavalli, P. Boutinaud, M. Bettinelli, P. Dorenbos, Lanthanide 4f-level location in $AVO_4:Ln^{3+}$ (A = La, Gd, Lu) crystals, J. Phys.: Condens. Matter. 21 (2009) 115503 - 115510. https://doi:10.1088/0953-8984/21/11/1155033.

[4] P. Yang, S. Huang, D. Kong, J. Lin, H. Fu, Luminescence functionalization of SBA-15 by $YVO_4:Eu^{3+}$ as a novel drug delivery system", Inorg Chem. 46 (2007) 3203-3211. https://doi.org/10.1021/ic0622959

[5] D. Sangla, M. Castaing, F. Balembois, P. Georges, Highly efficient Nd:YVO4 laser by direct in-band diode pumping at 914 nm, Optics Letters. 34 (2009) 2159-2161. https://doi.org/10.1364/OL.34.002159.

[6] J. Liu, Q. Yao, and Y. Li, Effects of downconversion luminescent film in dye-sensitized solar cells, Appl. Phys. Lett. 88 (2006) 173119(1-3). https://doi.org/10.1063/1.2198825.

[7] D.B. Barber, C.R. Pollock, L.L. Beecroft C.K. Ober, Amplification by optical composites, Optics Letters. 22 (1997) 1247-1249. https://doi.org/10.1364/OL.22.001247.

[8] Y.H. Chen, Y.C Huang, Actively Q-switched Nd:$YVO_4$ laser using an elecro-optic periodically poled lithium niobate crystal as a laser Q – switch. Optics Letters. 28 (2003) 1460-2. https://doi.org/10.1364/OL.28.001460.

[9] N.S, Lewis, D.G. Nocera, Powering the planet: Chemical challenges in solar energy utilization. Proc. Natl. Acad. Sci. U.S.A., 103 (2006) 15729-15735.

[10] B. van der Zwaan, A. Rabl, Prospects for PV: A Learning Curve Analysis. Sol. Energy, 74 (2003) 19-31. https://doi.org/10.1016/S0038-092X(03)00112-9

[11] Ke Xu , Miao Du, Lei Hao, Jing Mi, Qinghe Yu, Shijie Li, A review of high-temperature selective absorbing coatings for solar thermal applications, J. of Materiomics, 6 (2020) 167-182. https://doi.org/10.1016/j.jmat.2019.12.012

[12] P. Moraitis, R.E.I. Schropp, W.G.J.H.M. van Sark, Nanoparticles for Luminescent Solar Concentrators - A review. Optical Materials, 84 (2018) 636–645. https://doi.org/10.1016/j.optmat.2018.07.034

[13] B.M. Van der Ende, L. Aarts, A. Meijerink, Lanthanide ions as spectral converters for solar cells. Phys. Chem. Chem. Phys., 11 (2009) 11081-11095. https://doi.org/10.1039/B913877C

[14] X. Huang, S. Han, W. Huang, X. Liu, Enhancing solar cell efficiency: the search for luminescent materials as spectral converters. Chem. Soc. Rev. 42 (2013) 173-201. https://doi.org/10.1039/C2CS35288E

[15] H.Y. Lin, W.F. Chang, S.Y. Chu, Luminescence of $(Ca,Sr)_3(VO_4)_2:Pr^{3+},Eu^{3+}$phosphor for use in CuPc-based solar cells and white light-emitting diodes. J. Lumin., 133 (2013) 194-199. https://doi.org/10.1016/j.jlumin.2011.12.034

[16] V. Kumar, A.F. Khan, S. Chawla, Intense red-emitting multi-rare-earth doped nanoparticles of $YVO_4$ for spectrum conversion towards improved energy harvesting by solar cells, J. Phys. D: Appl. Phys. 46 (2013) 365101-9. http://npl.csircentral.net/id/eprint/3160.

[17] X. Yang, W. Zuo, F. Li, T. Li. Surfactant-Free and Controlled Synthesis of Hexagonal CeVO4 Nanoplates: Photocatalytic Activity and Superhydrophobic Property, Chemistry Open. 4 (2015) 288 – 294. https://doi.org/10.1002/open.201402163.

[18] V. Sivakumar, R. Suresh, K. Giribabu, V. Narayanan, $BiVO_4$ nanoparticles: Preparation, characterization and photocatalytic activity, Cogent Chemistry, 1 (2015) 1074647. https://doi.org/10.1080/23312009.2015.1074647

[19] Y. Zhu, Y. Wang, J. Zhu, D. Zhou, D. Qiu, W. Xu, X. Xu, Z. Lu, Plasmon multiwavelength-sensitized luminescence enhancement of highly transparent $Ag/YVO_4:Eu^{3+}$/PMMA film, J. Lumin. 200 (2018) 158-163. https://doi.org/10.1016/j.jlumin.2018.03.081.

[20] G. Liu, X. Duan, H. Li, H. Dong, Hydrothermal synthesis, characterization and optical properties of novel fishbone-like $LaVO_4:Eu^{3+}$ nanocrystals, Materials Chemistry and Physics. 115 (2009) 165–171. https://doi.org/10.1016/j.matchemphys.2008.11.043.

[21] T. Higuchi, Y. Hotta, Y. Hikita, S. Maruyama, Y. Hayamizu, H. Akiyama, H. Wadati, D.G. Hawthorn, T.Z. Regier, R.I.R. Blyth, G.A. Sawatzky, H.Y. Hwang, $LaVO_4$:Eu Phosphor films with enhanced Eu solubility, Appl. Phys. Let. 98 (2011) 071902(1-3). https://doi.org/10.1063/1.3554749.

[22] D. Errandonea, A.B. Garg Recent progress on the characterization of the high-pressure behaviour of $AVO_4$ orthovanadates, Progress in Materials Science. 97 (2018) 123 – 169. https://doi.org/10.1016/j.pmatsci.2018.04.004.

[23] C-J. Jia, L-D. Sun, Z-G. Yan, Yu-C. Pang, Shao-Zhe Lu, Chun-Hua Yan, Monazite and Zircon Type $LaVO_4$:Eu Nanocrystals – Synthesis, Luminescent Properties, and Spectroscopic Identification of the $Eu^{3+}$ Sites, Eur. J. Inorg. Chem. 18 (2010) 2626–2635. https://doi.org/10.1002/ejic.201000038.

[24] Chandresh Kumar Rastogi, Shilendra Kumar Sharma, Akshay Patel, G. Parthasarathy, Raj Ganesh S. Pala, Jitendra Kumar, Sri Sivakumar, Dopant Induced Stabilization of Metastable Zircon-Type Tetragonal $LaVO_4$. J. Phys. Chem. C. 121 (2017) 16501–16512. https://doi.org/10.1021/acs.jpcc.7b04508.

[25] H. Yuan, K. Wang, C. Wang, B. Zhou, K. Yang, J. Liu, B. Zou, Pressure-Induced Phase Transformations of Zircon-Type $LaVO_4$ Nanorods. J. Phys. Chem. C. 119 (2015) 8364–8372. https://doi.org/10.1021/acs.jpcc.5b01007.

[26] Jan W. Stouwdam, Mati Raudsepp, Frank C. J. M. van Veggel, Colloidal Nanoparticles of $Ln^{3+}$-Doped $LaVO_4$: Energy Transfer to Visible- and Near-Infrared-Emitting Lanthanide Ions, Langmuir. 21 (2005) 7003-7008. https://doi.org/10.1021/la0505162.

[27] B. Yan, X. Su, K. Zhou, In situ chemical co-precipatation composition of hybrid precursors to red $YVO_4$:$Eu^{3+}$ and green $LaPO_4$:$Tb^{3+}$ phosphors, Materials Research Bulletin. 41 (2006) 134–143. https://doi.org/10.1016/j.materresbull.2005.07.030.

[28] W.L. Fan, Y.X. Bu, X.Y. Song, S.X. Sun, X. Zhao, Selective synthesis and luminescent properties of monazite- and zircon-type $LaVO_4$:Ln (Ln = Eu, Sm, and Dy) nanocrystals, Cryst. Growth Des. 7 (2007) 2361-2366. https://doi.org/10.1021/cg060807o.

[29] G. Blasse, On the $Eu^{3+}$ Fluorescence of Mixed Metal Oxides. IV. The Photoluminescent Efficiency of $Eu^{3+}$ Activated Oxide, J. Chem. Phys. 45 (1966) 2356-2360. https://doi.org/10.1063/1.1727946.

[30] G. Blasse, B.C. Grabmaie, Luminescent Materials, Springer, New York, 1994. https://doi.org/10.1007/978-3-642-79017-1.

[31] L. Sun, X. Zhao, Ya. Li, Pan Li, H. Sun, X. Cheng, W. Fan, First-principles studies of electronic, optical, and vibrational properties of $LaVO_4$ polymorph, J. Appl. Phys. 108 (2010) 093519(10). https://doi.org/10.1063/1.3499308.

[32] A. Shyichuk, R.T. Moura, Jr., A.N. Carneiro Neto, M. Runowski, M.S. Zarad, Agata Szczeszak, Stefan Lis, O. L. Malt, Effects of Dopant Addition on Lattice and Luminescence Intensity Parameters of Eu(III)-Doped Lanthanum Orthovanadate, J. Phys. Chem. C. 120 (2016) 28497–28508. https://doi.org/10.1021/acs.jpcc.6b10778.

[33] J. Liu, Ya. Li, Synthesis and Self-Assembly of Luminescent $Ln^{3+}$- Doped $LaVO_4$ Uniform Nanocrystals, Adv. Mater. 19 (2007) 1118–1122. https://doi.org/10.1002/adma.200600336.

[34] C.J. Jia, L.D. Sun, L.P. You, X.C. Jiang, F. Luo, Y.C. Pang, C.H. Yan, Selective Synthesis of Monazite- and Zircon-Type LaVO4 Nanocrystals. J. Phys. Chem. B. 109 (2005) 3284-3290. https://doi.org/10.1021/jp045967u.

[35] Y. Oka, T. Yao, and N. Yamamoto, Hydrothermal Synthesis of Lanthanum Vanadates: Synthesis and Crystal Structures of Zircon-Type $LaVO_4$ and a New Compound $LaV_3O_9$, J. Solid State Chem. 152 (2000) 486 – 491. https://doi.org/10.1006/jssc.2000.8717.

[36] P. Gangwar, M. Pandey, S. Sivakumar, R.G.S. Pala, G. Parthasarathy, Increased Loading of $Eu^{3+}$ Ions in Monazite $LaVO_4$ Nanocrystals via Pressure-Driven Phase Transitions, Cryst. Growth Res. 13 (2013) 2344-2349. https://doi.org/10.1021/cg3018908.

[37] F. He, P. Yang, D. Wang, N. Niu, S. Gai, X. Li, M. Zhang, Hydrothermal Synthesis, Dimension Evolution and Luminescence Properties of Tetragonal $LaVO_4$: Ln (Ln= $Eu^{3+}$, $Dy^{3+}$, $Sm^{3+}$) Nanocrystals, Dalton Trans. 40 (2011)11023-11030. https://doi.org/10.1039/c1dt11157d.

[38] R. van Deun, M. D'hooge, A. Savic, I.V. Driessche, K.V. Hecke, A.M. Kaczmarek, Influence of $Y^{3+}$, $Gd^{3+}$, and $Lu^{3+}$ co-doping on the phase and luminescence properties of monoclinic Eu:$LaVO_4$ particles, Dalton Trans. 44 (2015) 18418-18426. https://doi.org/10.1039/c5dt03147h

[39] J.H. Kang, W.B. Im, D. Chin Lee, J.Y. Kim, D.Y. Jeon, Y.C. Kang, K.Y. Jung, Correlation of photoluminescence of (Y, Ln)$VO_4$:$Eu^{3+}$ (Ln=Gd and La) phosphors with their crystal structures, Solid State Comm. 133 (2005) 651-656. https://doi.org/10.1016/j.ssc.2005.01.004.

[40] R. Lisiecki, W. Ryba-Romanowski, E. Cavalli, M. Bettinelli, Optical spectroscopy of $Er^{3+}$-doped $LaVO_4$ crystal, J. Lumin. 130 (2010) 131–136. https://doi.org/10.1016/j.jlumin.2009.07.037.

[41] R. Lisiecki, G. Dominiak-Dzik, P. Solarz, A. Strzęp, W. Ryba-Romanowski, T. Łukasiewicz, Spectroscopic characterisation of Er-doped LuVO4 single crystals, Appl. Phys. B. 101 (2010) 791–800. https://doi.org/10.1007/s00340-010-4212-6.

[42] R. Okram, G. Phaomei, N. Rajmuhon Singh, Water driven enhanced photoluminescence of Ln (= $Dy^{3+}$, $Sm^{3+}$) doped $LaVO_4$ nanoparticles and effect of $Ba^{2+}$ co-doping, Materials Science and Engineering B. 178 (2013) 409–416. http://dx.doi.org/10.1016%2Fj.mseb.2013.01.007.

[43] O. Chukova, S.A. Nedilko, M. Androulidaki, S.G. Nedilko, A. Slepets, . P.T. Besa, T. Voitenko, Structure, Morphology and Spectroscopy Characterization of the $La_{1-X-Y}Er_{x/2}Eu_{x/2}Ca_yVO_4$ Sol-Gel Nanoparticles, in: 2019 IEEE 39th International Conference on Electronics and Nanotechnology ELNANO Proceedings. (2019) 261-264. DOI: 10.1109/ELNANO.2019.8783459.

[44] O.V. Chukova, S.G. Nedilko, A.A. Slepets, S.A. Nedilko, T.A. Voitenko, Synthesis and Investigation of La,Ca -Doped $EuVO_4$ Nanoparticles with Enhanced Excitation by Near Violet Light, Phys. Stat. Solidi A. 215 (2018) 1700894-7. https://doi.org/10.1002/pssa.201700894.

[45] O. Chukova, S.A. Nedilko, S.G. Nedilko, A. Slepets, T. Voitenko, M. Androulidaki, A. Papadopoulos, E.I. Stratakis, Structure, morphology, and spectroscopy studies of $La_{1-x}RE_xVO_4$ Nanoparticles Synthesized by Various Methods, in: O. Fesenko, A. Yatsenko (Eds.), Springer Proceedings in Physics. 221 (2019) 211-241. https://doi.org/10.1007/978-3-030-17759-1_15.

[46] O. Chukova, S.A. Nedilko, S.G. Nedilko, A. A. Papadopoulos, A. Slepets, E.I. Stratakis, T. Voitenko, Structure and spectroscopy characterization of $La_{1-x}Sm_xVO_4$ luminescent nanoparticles synthesized co-precipitation and sol-gel methods, Optical Materials. 95 (2019) 109248. https://doi.org/10.1016/j.optmat.2019.109248.

[47] W. Paszkowicz, J. López-Solano, P. Piszora, B. Bojanowski, A. Mujica, A. Muñoz, Y. Cerenius, S. Carlson, H. Dąbkowska, Equation of state and electronic properties of $EuVO_4$: A high-pressure experimental and computational study, Journal of Alloys and Compounds, 648 (2015) 1005-1016. https://doi.org/10.1016/j.jallcom.2015.06.211

[48] Manoj Kumar Mahata, Surya Prakash Tiwari, Shriparna Mukherjee, Kaushal Kumar, Vineet Kumar Rai, $YVO_4:Er^{3+}/Yb^{3+}$ phosphor for multifunctional applications, J. Opt. Soc. Am. B. 31 (2014) 1814-1821. https://doi.org/10.1364/JOSAB.31.001814.

[49] P. Parhi, V. Manivannan, Novel microwave initiated solid-state metathesis synthesis and characterization of lanthanide phosphates and vanadates, $LMO_4$ (L = Y, La and M = V, P), Solid State Sciences. 10 (2008) 1012 – 1019. https://doi.org/10.1016/j.solidstatesciences.2007.11.038.

[50] T. Nakajima, M. Isobe, T. Tsuchiya, Y. Ueda, T. Manabe, Correlation between Luminescence Quantum Efficiency and Structural Properties of Vanadate Phosphors with Chained, Dimerized, and Isolated $VO_4$ Tetrahedra, J. Phys. Chem. C. 114 (2010) 5160–5167. https://doi.org/10.1021/jp910884c.

[51] O. Ermakova, J. López-Solano, R. Minikayev, S. Carlson, A. Kamińska, M. Głowacki, M. Berkowski, A. Mujica, A. Muñoz, W. Paszkowicz, A combined study of the equation of state of monazite-type lanthanum orthovanadate using in situ high-pressure diffraction and ab initio calculations. Acta Crystallogr. B, 70 (2014) 533-538. https://doi.org/10.1107/S2052520614010816.

[52] A.A. Ansari, M. Alam, J.P. Labis, S.A. Alrokayan, G. Shafi, T.N. Hasan, N.A. Syed, A.A. Alshatwi, Luminescent mesoporous $LaVO_4:Eu^{3+}$ core-shell nanoparticles: synthesis, characterization, biocompatibility and their cytotoxici, J. Mater. Chem. 21 (2011) 19310–19316. https://doi.org/10.1039/c1jm12871j.

[53] R. Van Deun, K. Binnemans, C. Görller-Walrand, J.L. Adam, Optical properties of $Eu^{3+}$-doped fluorophosphate glasses, J. Phys.: Condens. Matter 10 (1998) 7231–7241. https://doi.org/10.1088/0953-8984/10/32/014.

[54] Z. Wang, J. Zhong, H. Liang, J. Wang, Luminescence properties of lutetium based red-emitting phosphor $NaLu(WO_4)_2:Eu^{3+}$, Optical Materials Express 3 (2013) 418-425. https://doi.org/10.1364/OME.3.000418.

[55] O. Chukova, S. Nedilko, V. Scherbatskyi, Luminescent spectroscopy and structure of centers of the impurity $Eu^{3+}$ ions in lead tungstate crystals, J. Lumin. 130 (2010) 1805-1812. https://doi.org/10.1016/j.jlumin.2010.04.014.

[56] R.D. Shannon, Revised effective ionic radii and systematic studies of interatomic distances in halides and chalcogenides, Acta Crystallogr. A. 32, (1976) 751–767. https://doi.org/10.1107/S0567739476001551.

[57] M. Huse, T. Norby, R. Haugsrud, Proton Conductivity in Acceptor-Doped $LaVO_4$. J. of The Electrochem. Soc. 158 (2011) B857-B865. https://doi.org/10.1149/1.3594145.

[58] R. Jablonski, S.M. Kaczmarek, M. Swirkowicz, T. Lukasiewi, Electron spin resonance and optical measurements of yttrium ortho-vanadate, doped with Nd ions, J. of Alloys and Compounds. 300–301 (2000) 310–315. https://doi.org/10.1016/S0925-8388(99)00730-6.

[59] N.Y. Garces, K.T. Stevens, G.K. Foundos, L.E. Halliburton, Electron paramagnetic resonance and optical absorption study of $V^{4+}$ centres in $YVO_4$ crystals, J. of Phys.: Condensed Matter. 16 (2004) 7095. https://doi.org/10.1088/0953-8984/16/39/040.

[60] V. Chornii, O. Chukova, S.G. Nedilko, S.A. Nedilko, T. Voitenko, Enhancement of emission intensity of $LaVO_4$:$RE^{3+}$ luminescent solar light absorbers, Phys. Stat. Solidi C. 13 (2016) 40–46. https://doi.org/10.1002/pssc.201510116.

[61] L. Yang, G. Li, W. Hu, M. Zhao, L. Sun, J. Zheng, T. Yan, L. Li, Control Over the Crystallinity and Defect Chemistry of $YVO_4$ Nanocrystals for Optimum Photocatalytic Property, Eur. J. Inorg. Chem. 14 (2011) 2211-2220. https://doi.org/10.1002/ejic.201001341.

[62] S.G. Nedilko, V. Chornii, O. Chukova, V. Degoda, K. Bychkov, K. Terebilenko, M. Slobodyanik, Luminescence properties of the new complex La,Bi$VO_4$: Mo,Eu compounds as materials for down-shifting of VUV–UV radiation, Radiation Measurements. 90 (2016) 282-286. https://doi.org/10.1016/j.radmeas.2016.02.027.

[63] S.W. Park, H.K. Yang, J.W. Chung, Y. Chen, B.K. Moon, B.C. Cho, J.H. Jeon, J.H. Kim, Photoluminescent properties of $LaVO_4$ $Eu^{3+}$ by structural transformation, Physica B. 405 (2010) 4040–4044. https://doi.org/10.1016/j.physb.2010.06.052.